\documentclass[final]{aipproc}
\layoutstyle{6x9}
\begin{document}
\title{Mesonic states in the generalised Nambu-Jona-Lasinio theories}

\author{A. V. Nefediev}{address={Institute of Theoretical and Experimental Physics, 117218,
B.Cheremushkinskaya 25, Moscow, Russia}}

\author{J. E. F. T. Ribeiro}{address={Centro de F\'\i sica das Interac\c c\~oes Fundamentais
(CFIF), Departamento de F\'\i sica, Instituto Superior T\'ecnico, Av.
Rovisco Pais, P-1049-001 Lisboa, Portugal}}

\newcommand{\be}{\begin{equation}}
\newcommand{\bea}{\begin{eqnarray}}
\newcommand{\ee}{\end{equation}}
\newcommand{\eea}{\end{eqnarray}}
\newcommand{\ds}{\displaystyle}
\newcommand{\low}[1]{\raisebox{-1mm}{$#1$}}
\newcommand{\loww}[1]{\raisebox{-1.5mm}{$#1$}}
\newcommand{\lmn}{\mathop{\sim}\limits_{n\gg 1}}
\newcommand{\vpint}{\int\makebox[0mm][r]{\bf --\hspace*{0.13cm}}}
\newcommand{\too}{\mathop{\to}\limits_{N_C\to\infty}}
\newcommand{\vp}{\varphi}

\begin{abstract}
For any Nambu-Jona-Lasinio model of QCD with arbitrary nonlocal,
instantaneous, quark current-current confining kernels, we use a
generalised Bogoliubov technique to go beyond BCS level (in the large-$N_C$
limit) so as to explicitly build quark-antiquark compound operators for
creating/annihilating mesons. In the Hamiltonian approach, the mesonic
bound-state equations appear (from the generalised Bogoliubov
transformation) as mass-gap-like equations which, in turn, ensure the
absence, in the Hamiltonian, of mesonic Bogoliubov anomalous terms. We go
further to demonstrate the one-to-one correspondence between Hamiltonian
and Bethe-Salpeter approaches to non-local NJL-type models for QCD and
give the corresponding "dictionary" necessary to "tran\-sla\-te" the
amplitudes built using the graphical Feynman rules to the terms of the
Hamiltonian, and vice versa. We comment on the problem of multiple vacua
existence in such type of models and argue that mesonic states in the
theory should be prescribed to have an extra index --- the index of
the replica in which they are created. Then the completely diagonalised
Hamiltonian should contain a sum over this new index. The method is proved
to be general and valid for any instantaneous quark kernel. 
\end{abstract}

\maketitle

We study generalised Nambu-Jona-Lasinio models which are expected to mimic the most important 
low-energy properties of QCD \cite{NJL,NJL2}. The Hamiltonian reads:
\be
\hat{H}=\int d^3 x\bar{\psi}(\vec{x},t)\left(-i\vec{\gamma}\cdot
\vec{\bigtriangledown}+m\right)\psi(\vec{x},t)+ \frac12\int d^3
xd^3y\;J^a_\mu(\vec{x},t)K^{ab}_{\mu\nu}(\vec{x}-\vec{y})J^b_\nu(\vec{y},t),
\label{H}
\ee
and contains the interaction of the quark currents
$J_{\mu}^a(\vec{x},t)=\bar{\psi}(\vec{x},t)\gamma_\mu\frac{\lambda^a}{2}
\psi(\vec{x},t)$, parameterised through the instantaneous quark
kernel, $K^{ab}_{\mu\nu}(\vec{x}-\vec{y})=g_{\mu 0}g_{\nu 0}\delta^{ab}
V_0(|\vec{x}-\vec{y}|)$, with a power-like confining potential $V_0(|\vec{x}|)=K_0^{\alpha+1}|\vec{x}|^{\alpha}$.
The standard approach to the theories (\ref{H}) is the Bogoliubov-Valatin transformation defined by the chiral angle
$\vp_p$ \cite{NJL2},
\be 
\left\{
\begin{array}{rcl}
u(\vec{p})&=&\frac{1}{\sqrt{2}}\left[\sqrt{1+\sin\vp_p}+
\sqrt{1-\sin\vp_p}\;(\vec{\alpha}\hat{\vec{p}})\right]u(0),\\
v(-\vec{p})&=&\frac{1}{\sqrt{2}}\left[\sqrt{1+\sin\vp_p}-
\sqrt{1-\sin\vp_p}\;(\vec{\alpha}\hat{\vec{p}})\right]v(0).
\end{array}
\right.
\label{uandv}
\ee
Anomalous terms in the Hamiltonian vanish if $\vp_p$ obeys the mass-gap equation,
\be
m\cos\vp_p-p\sin\vp_p=\frac{C_F}{2}\int\frac{d^3k}{(2\pi)^3}V_0(\vec{p}-\vec{k})
\left[\sin\vp_k\cos\vp_p-(\hat{\vec{p}}\hat{\vec{k}})\cos\vp_k\sin\vp_p\right].
\label{mg}
\ee
Eq.~(\ref{mg}) is subject to numerical studies. As soon as a nontrivial solution $\vp_0(p)$ to the mass-gap equation is 
built, it defines the vacuum of the theory with spontaneously broken chiral symmetry, which is energetically
preferable as compared to the unbroken phase. The terms in (\ref{H}) quartic in quark operators contain
the suppressing factor of $1/\sqrt{N_C}$ and thus the Hamiltonian of the theory is diagonalised in the quark
sector (BCS level). 

To proceed beyond BCS level and reformulate the theory in terms of colourless mesonic
states, we notice that only operators creating/annihilating $q\bar q$ pairs are allowed:
\be
\hat{M}_{ss'}(\vec{p},\vec{p})=\frac{\ds
1}{\ds\sqrt{N_C}}\hat{d}_{\alpha s}(-\vec{p})\hat{b}_{\alpha
s'}(\vec{p})=\sum_\nu[\kappa_\nu(\hat{\vec{p}})]_{ss'}\sum_n[\hat{m}_{n\nu}\vp_{n\nu}^+(p)+\hat{m}_{n\nu}^\dagger\vp_{n\nu}^-(p)],
\label{M}
\ee
all other operators, like $\hat{b}^\dagger \hat{b}$ and $\hat{d}^\dagger \hat{d}$, being suppressed by $N_C$ \cite{2d,replica1}. 
In (\ref{M}) we
consider the $q\bar q$ pair at rest and separate the spin-angular and the radial parts of the operator
$\hat{M}$. The complete set $\{\kappa_\nu\}$ is chosen to be the $J^{PC}$ set of states, which are known to
diagonalise the Hamiltonian of strong interactions. We also perform a second, generalised, Bogoliubov-like
transformation and introduced the mesonic creation/annihilation operators $\hat{m}_{n\nu}^\dagger/\hat{m}_{n\nu}$. 
The Bogoliubov amplitudes $\vp^\pm_{n\nu}$ obey the normalisation condition which follows immediately from
the commutation relation for the mesonic operators:
\be
[\hat{m}_{n\nu},\;\hat{m}_{m\nu}^\dagger]=\ds\int\frac{p^2dp}{(2\pi)^3}\left[\vp_{n\nu}^{+}(p)\vp_{m\nu}^{+}(p)-
\vp_{n\nu}^{-}(p)\vp_{m\nu}^{-}(p)\right]=\delta_{nm}.
\label{normgen}
\ee

Meanwhile, $\vp^\pm_{n\nu}$ also play the role of the two components of the mesonic wave function responsible
for the forward and backward in time motion of the $q\bar q$ pair in the meson. They are required to be
solutions of an eigenvalue problem --- the bound-state equation; $M_{n\nu}$ being the mass of the
corresponding meson:
\be
\left\{
\begin{array}{l}
[2E_p-M_{n\nu}]\vp_{n\nu}^+(p)=\ds\int\frac{\ds q^2dq}{\ds (2\pi)^3}
[T^{++}_\nu(p,q)\vp_{n\nu}^+(q)+T^{+-}_\nu(p,q)\vp_{n\nu}^-(q)]\\[0cm]
[2E_p+M_{n\nu}]\vp_{n\nu}^-(p)=\ds\int\frac{\ds q^2dq}{\ds (2\pi)^3}
[T^{-+}_\nu(p,q)\vp_{n\nu}^+(q)+T^{--}_\nu(p,q)\vp_{n\nu}^-(q)].
\end{array}
\right.
\label{bsgen}
\ee

Eq.~(\ref{bsgen}) can be alternatively derived using the Bethe--Salpeter approach to mesonic states \cite{NJL2}, and 
thus a close connection between the Bethe--Salpeter and Hamiltonian approaches to the theory can be
established, including graphical rules, which allow one to build the $T$-amplitudes in Eq.~(\ref{bsgen}) using
Feynman-like diagrams \cite{replica1}. The Hamiltonian (\ref{H}) takes now a diagonal form in terms of
mesonic operators,
\be
\hat{\cal H}=\sum_{n,\nu}M_{n\nu}m^\dagger_{n\nu}m_{n\nu},
\label{hdgen}
\ee
with corrections to the leading regime (\ref{hdgen}) suppressed by $N_C$. The first correction, of
order $1/\sqrt{N_C}$, is responsible for mesonic decays \cite{2d}. Notice that, as soon as the mass-gap
equation is solved, no new information, at least in the leading order in $N_C$, is needed to proceed 
beyond the BCS level and to introduce mesonic states. Numerical analysis of the mass-gap equation (\ref{mg})
for various confining potentials $V(r)$ demonstrated existence of extra, "excited", solutions --- the
vacuum replicas \cite{NJL2,replica2}. The existence of an infinite tower of such replicas for power-like
confining potentials $V(r)\propto r^\alpha$ was proved analytically and verified numerically for all
$\alpha$'s from the allowed region $0\leq\alpha\leq 2$, as well as for D=2 and D=4 (D is the
dimensionality of the space-time) \cite{replica3}. It was argued in \cite{replica3,replica1} that the
appearance of such replicas is a consequence of a very peculiar behaviour of the dressed quark dispersive law
$E_p$ in the infrared region and, therefore, is closely related to chiral symmetry breaking.
A similar conclusion was also made in a different approach in \cite{replica4}. It was demonstrated in
\cite{replica1} that, with the proper definition of the chiral angle, one encounters no problem with the 
imaginary mass of the pion in the replica vacua --- all mesons build in replicas being normal hadronic
states, but with a heavier mass as compared to mesons created in the ground-state vacuum. It is quite natural
then to require that the information about vacuum replicas is transfered beyond the BCS level, the full
Hamiltonian should contain the sum over the index $\cal N$ numerating the replicas,
\be
\hat{\cal H}=\sum_{n,\nu,{\cal N}}M_{n\nu{\cal N}}m^\dagger_{n\nu{\cal N}}m_{n\nu{\cal N}}.
\label{hdgen2}
\ee

In conclusion let us notice that the analysis of the generalised Nambu-Jona-Lasinio theories performed above is
insensitive to the dimensionality of the space-time, the Lorentz nature of confinement, and its 
explicit form. We argue that the existence of replicas is a rule rather than an exception for any confining 
quark kernel and we believe it is not an artifact of the instantaneous interquark interaction used in our
analysis in order to bypass the problem of the relative time. An approach
to replicas, as to local excitations, was suggested in \cite{replica5} and an effective diagrammatic technique was derived in order to take
into account the effect of replicas in hadronic reactions. Another important ingredient of the theory of
vacuum replicas is a mechanism of excitation of replicas as global objects. This work is in progress now and
will be reported elsewhere.
\bigskip

One of the authors (A.V.N.) would like to acknowledge discussions with P. Bicudo and Yu. S.
Kalashnikova as well as the financial support of the grant NS-1774.2003.2, 
as well as of the Federal Programme of the Russian Ministry of Industry, Science, and
Technology No 40.052.1.1.1112.


\begin{thebibliography}{99}
\expandafter\ifx\csname natexlab\endcsname\relax\def\natexlab#1{#1}\fi
\providecommand{\enquote}[1]{``#1''}
\expandafter\ifx\csname url\endcsname\relax
  \def\url#1{\texttt{#1}}\fi
\expandafter\ifx\csname urlprefix\endcsname\relax\def\urlprefix{URL }\fi


\bibitem{NJL} Nambu Y., Jona-Lasinio G., {\it Phys. Rev.}, {\bf 122}, 345 (1961).
\bibitem{NJL2} Amer A., Le Yaouanc A., Oliver L., Pene O., and 
Raynal J.-C., {\it Phys. Rev. Lett.}, {\bf 50}, 87 (1983); Le Yaouanc A., Oliver L.,
Pene O., and Raynal J.-C., {\it Phys. Lett.}, {\bf 134B}, 249 (1984); {\it Phys.
Rev.}, {\bf D29}, 1233 (1984); Bicudo P. and Ribeiro J. E., {\it Phys. Rev.}, {\bf D42},
1611 (1990); 1625 (1990); 1635 (1990); Bicudo P., {\it Phys. Rev. Lett.}, {\bf 72}, 1600 (1994); 
Bicudo P., {\it Phys. Rev.}, {\bf C60}, 035209 (1999).
\bibitem{2d} Kalashnikova Yu. S. and Nefediev A. V., {\it Phys. Atom. Nucl.}, {\bf 62}, 323 (1999)
{\it Phys. Usp.}, {\bf 45}, 347 (2002); Kalashnikova Yu. S.,
Nefediev A. V., and Volodin A. V., {\it Phys. Atom. Nucl.}, {\bf 63}, 1623 (2000).
\bibitem{replica1} Nefediev A. V. and Ribeiro J. E. F. T., {\it Phys. Rev. D}, in press.
\bibitem{replica2} Bicudo P. J. A., Nefediev A. V., and Ribeiro J. E. F. T., {\it Phys. Rev.}, 
{\bf D65}, 085026 (2002).
\bibitem{replica3} Bicudo P. J. A. and Nefediev A. V., {\it Phys. Rev.}, {\bf D68}, 065021 (2003); 
{\it Phys. Lett.} {\bf 573B}, 131 (2003).
\bibitem{replica4} Osipov A. A. and Hiller B., {\it Phys. Lett.}, {\bf 539B}, 76 (2002).
\bibitem{replica5} Nefediev A. V. and Ribeiro J. E. F. T., {\it Phys. Rev.}, {\bf D67}, 034028 (2003).
\end{thebibliography}
\end{document}